\documentclass[12pt]{article}
\usepackage{amsmath,amsfonts,epsfig,mathrsfs,todonotes,yfonts}
\usepackage{calligra}  
\usepackage{dsfont}
\usepackage{stmaryrd}

\usepackage{hyperref}
 \hypersetup{
     colorlinks=true,
     linkcolor=black,
     filecolor=back,
     citecolor=blue,      
     urlcolor=cyan,
     }
\usepackage[most]{tcolorbox}

\numberwithin{equation}{section}

\newcommand{\al}{\alpha}

\newcommand{\one}{{\mathds{1}}}

\newcommand{\nn}{\nonumber}

\newcommand{\ber}{\begin{eqnarray}}
\newcommand{\eer}[1]{\label{#1}\end{eqnarray}}
\newcommand{\eero}{\end{eqnarray}}

\newcommand{\half}{{\textstyle{\frac12}}}

\newcommand{\bbD}[1]{\mathbb{D}_{#1}}
\newcommand{\bbDB}[1]{\bar{\mathbb{D}}_{#1}}
\def\+{{+\!\!\!+}} 
\def\pp{\mbox{\tiny${}_{\stackrel\+ =}$}}
\newcommand{\pa}[1]{\partial_{#1}}

\newcommand{\re}[1] {(\ref{#1})}

 \DeclareMathAlphabet{\mathcalligra}{T1}{calligra}{m}{n}
\DeclareFontShape{T1}{calligra}{m}{n}{<->s*[2.2]callig15}{}

\newcommand{\auth}
{\large Ulf Lindstr\"om ${}^{a,b}$\footnote{ Leverhulme Visiting Professor. 
email: ulf.lindstrom@physics.uu.se}}
\usepackage{tikz}

\newcommand*\circled[1]{\tikz[baseline=(char.base)]{%
            \node[shape=circle,fill=blue!20,draw,inner sep=2pt] (char) {#1};}}

\usepackage{enumitem}

\begin{document}

\hfill UUITP-24/23 
%\hfill\today 
\vspace{1cm}
                    \begin{center}
{\Large{\bf Yano $F$ structures and extended Supersymmetry in a BiLP.}}\\
\vspace{0.75cm}
\auth
\end{center}
\vspace{.5cm}
\centerline{${}^a${\it \small Theoretical Physics, Imperial College London,}}
\centerline{{\it \small Prince Consort Road, London SW7 2AZ, UK}}

\vspace{.5cm}
\centerline{${}^b${\it \small Department of Physics and Astronomy, Theoretical Physics, Uppsala University}}
\centerline{{\it \small SE-751 20 Uppsala, Sweden }}
\bigskip

{\bf Abstract}: \vspace{0,5cm} 

In this paper we look at a sigma model based on the $(4,4)$ twisted chiral multiplet \cite{Gates:1984nk}. It admits two geometric descriptions: The ususal  biquaternionic geometry on the tangent space and a new geometry involving two Yano F-structures on a doubled  tangent space. This analysis sheds light on the recent discussion of  semichiral models in \cite{Lindstrom:2022nto} and on an upcoming discussion of complex linears.
\eject

\tableofcontents
\section{Introduction}
In a recent article\footnote{Partly drawing on \cite{Goteman:2009ye}.} \cite{Lindstrom:2022nto} on  $(2,2)$ symplectic sigma models,
it is shown that additional supersymmetries imply the existence of two integrable Yano $F$ structures. 

The analysis deals with a general symplectic model and would benefit from an explicit example. 
In this note we illustrate the general procedure in the transparent case of the $(4,4)$ twisted 
chiral multiplet \cite{Gates:1984nk}. Unlike in the symplectic model there are no auxiliary fields to be integrated out so the relation between $(2,2)$ and $(1,1)$ formulations is very direct. Nevertheless the two geometries found in the symplectic model are found here too.

\section{Background}
\subsection{ $(1,1)$ Geometry of sigma models } 
\label{ex1}
Additional symmetries of a $(1,1)$ supersymmetric non linear sigma model in two dimensions are associated with additional complex structures $J^i$. The transformations follow the pattern
\ber
\delta^i\varphi=\epsilon_i^\pm J^i\bbD{\pm}\varphi~.
\eer{} 
When the extra symmetry is $(4,4)$, the corresponding complex structures form an $SU(2)$ algebra and are covariantly constant with respect to the Levi Civita connection (hyperk\"ahler) or the Levi-Civita connection plus torsion (bi hermitian). The metric is hermitian with respect to all complex structures.

\subsection{ $(2,2)$ Geometry of pre-sigma models } 
The $(2,2)$ algebra is 
%This section corresponds to item one in the list subsectionre{enu}.
\ber
\{\bbD{\pm},\bbDB{\pm}\}=2i\pa{\pp}~.
\eer{}
The  symplectic model discussed in \cite{Lindstrom:2022nto} is based on left $\ell$ and right $\mathfrak{r}$ semichiral superfields
\ber
\bbDB{+}\ell=0~,~~~\bbDB{-}\mathfrak{r}=0~.
\eer{}
These $(2,2)$ fields contain  auxiliary spinorial $(1,1)$ superfields  which have to be integrated out before the sigma model structure can be seen. So the $(2,2)$ formulation is not a sigma model and we don't expect it to carry the usual geometry. Instead it was found in \cite{Lindstrom:2022nto} that an ansatz 
\ber
\delta\varphi=U^{(\al)}\bar\epsilon^\al\bbDB{\al}\varphi+V^{(\al)}\epsilon^\al\bbD{\al}\varphi~,~~~~\al\in(+,-)
\eer{TFS}
leads to the existence of two Yano $F$-structures ${\cal F}_{(\al)}$ on the doubled tangent bundle.

The reduction of the model to $(1,1)$ superspace, integrating out the spinorial auxiliaries, leads to the usual  geometry described in Sec. \ref{ex1}, as it should.

\subsection{Strategy}

The procedure  in \cite{Lindstrom:2022nto} can be summarised as follows:
\begin{enumerate}[label=\protect\circled{\arabic*}]\label{enu}
\item In $(2,2)$ superspace, consider a sigma model based on $(2,2)$ superfields $\varphi$ and make the most general ansatz for additional susy transformations in terms of $\varphi$, preserving all their chirality conditions.
\item Use closure of the susy algebra and invariance of the action to constrain the transformations.
\item Identify the geometric objects that result from step 2. E.g.  the Yano F-stuctures  for the symplectic model in \cite{Lindstrom:2022nto}.
\item Reduce the resulting transformations and structures to $(1,1)$ superspace where the geometry corresponding to extended susy is wellknown.
\end{enumerate}
\bigskip

Below we check these steps for the  twisted chiral $(4,4)$ multiplet \cite{Gates:1984nk} where all structure is known.
%Points 1 and 2 in the list above 
\section{The $N=4$ twisted chiral multiplet}
In $(2,2)$ superspace the $N=4$ twisted chiral multiplet is given in terms of
{chiral superfields $\phi$ and {twisted chiral superfields} $\chi$
\ber\nn
&&\bbDB{\pm}\phi=0~\\[1mm]
&&\bbDB{+}\chi=\bbD{-}\chi=0~.
\eer{}
A model based on these has
an action
\ber
S=\int d^2xd^2\theta d^2\bar\theta K(\phi, \bar \phi, \chi,\bar \chi)~.
\eer{Achirtchir1}
Reduction to  $(1,1)$ superspace reveals that the target space geometry has two commuting complex structures, a local product structure and a metric that is hermitean wrt both complex structures: A Bihermitean Local Product (BiLP) geometry \cite{Gates:1984nk}, \cite{Lindstrom:2007qf}. This model has $(4,4)$ off-shell supersymmetry when the $ (2,2)$ fields sit in the  $(4,4)$ twisted chiral multiplet,  \cite{Gates:1984nk}
 provided that
\ber
&&K,_{\phi\bar\phi}+K,_{\chi \bar\chi}=0%\\[1mm]~,
%&&K,_{\phi^i \bar\phi^j}-K,_{\phi^j \bar\phi^i}=0~.
\eer{CNDS}
and that the complex structures are covariantly constant wrt  connections with torsion,
as shown in  \cite{Gates:1984nk}. 

A description in terms of transformations of the $ (2,2)$ fields instead
leads to Yano $F$ structures, which we now discuss.

\subsection{Transformations}

To simplify the description we consider the case of one chiral $\phi$ and one twisted chiral multiplet $\chi$ transforming under the extra susy as \cite{Gates:1984nk}
\ber\nn
&&\delta  \phi=\bar\epsilon^+\bbDB{+}{\bar\chi}+\bar\epsilon^-\bbDB{-}\chi\\[1mm]\nn
&&\delta {\bar\phi}=\epsilon^+\bbD{+}{\chi}+\epsilon^-\bbD{-}{\bar\chi}\\[1mm]\nn
&&\delta {\chi}=-\bar\epsilon^+\bbDB{+}{\bar\phi}-\epsilon^-\bbD{-}{\phi}\\[1mm]
&&\delta {\bar\chi}=-\epsilon^+\bbD{+}{\phi}-\bar\epsilon^-\bbDB{-}{\bar\phi}~.
\eer{4tfs}
Defining 
\ber
\varphi:=\left(\begin{array}{c} \phi\\\bar\phi\\\chi\\\bar\chi\end{array}\right)~,
\eer{}
we write \re{4tfs} as
\ber
\delta\varphi=U^{(\pm)}\bar\epsilon^\pm\bbDB{\pm}\varphi+V^{(\pm)}\epsilon^\pm\bbD{\pm}\varphi
\eer{TFS}
where $V=\bar U$ suitably rearranged, and
\ber
U^{(+)}=\left(\begin{array}{cccc} 
0&0&0&1\\
0&0&0&0\\
0&-1&0&0\\
0&0&0&0
\end{array} \right)~~~,~~~V^{(+)}=\left(\begin{array}{cccc} 
0&0&0&0\\
0&0&1&0\\
0&0&0&0\\
-1&0&0&0
\end{array} \right)~,
\eer{UVdefs}
and similarly for $U^{(-)}$ and $V^{(-)}$. We find that $U^{(+)}V^{(+)}=-diag(1,0,1,0)$, 
$V^{(+)}U^{(+)}=-diag(0,1,0,1)$
and that $U^{(-)}V^{(-)}=-diag(1,0,0,1)$  and  $V^{(-)}U^{(-)}=-diag(0,1,1,0)$ . 

{\em Closure of the extended susy algebra} leads to the Nijenhuis tensors for $U^{
\al}$ and $V^{\al}$ vanishing.
\subsection{Geometry}
The above results  allows us to define two Yano $F$-structures ${\cal F}_{(\pm)}$ \cite{Yano:1961} \cite{IshiharaYano}  acting on the doubled\footnote{Whitney sum} tangent space $T\oplus T$:
\ber
{\cal F}_{(\al)}:=\left(\begin{array}{cc} 
0&U^{(\al)}\\
V^{(\al)}&0
\end{array} \right)~,~~~{\cal F}^3_{(\al)}+{\cal F}_{(\al)}=0~.
\eer{}
An $F$ structure defines projection operators
\ber\nn
&&l:=-{\cal F}^2=-\left( \begin{array}{cc}
UV&0\cr 
0&VU\cr
\end{array}\right) \makebox{and} 
~m:=\one+{\cal F}^2=\left( \begin{array}{cccc}
\one+UV&0\cr
0&\one+VU\cr
\end{array}\right)\\
&&~
\eer{}
that obey
\ber\nn
&&	l   + m   = \one,\quad
	l^2   = l  , \quad m^2   = m  ~,\quad l   m =ml  = 0 \\[1mm]
&& 
	{\cal F}   l   = l   {\cal F}   = {\cal F}  , \quad m  {\cal F}   ={\cal F}   m  = 0~.
\eer{fund1}

The subspace corresponding to the projector  $m_{(+)}$ is  $(\phi,\chi)\oplus (\bar\phi,\bar\chi)$ while the complement 
$(\bar\phi,\bar\chi)\oplus (\phi,\chi)$  corresponds to the projector $l_{(+)}$;  $m_{(-)}$ projects onto  $(\phi,\bar\chi)\oplus (\bar\phi,\chi)$ with the complement 
$(\bar\phi,\chi)\oplus (\phi,\bar\chi)$.

In the special case when all entries in the $U$ and $V$ matrices are derivatives of a function,\footnote{Certainly true in the present $(4,4)$ twisted chiral  case.}
the condition from 
 {\em invariance of the action} \re{Achirtchir1}
 for the $\bar\epsilon^{\al} $ transformations in \re{TFS} reads:
\begin{equation}
\label{hermiticity}
0=\left(K,_{i}U^{(\al)i}{}_{[j}\right){}_{k]}=K,_{i[j}U^{(\al)i}{}_{k]}~, 
\end{equation}
The conditions for invariance under $\epsilon^{\al} $ transformations follow by complex conjugation, replacing $U\to V$.
Using the explicit form of $U$ and $V$ in \re{UVdefs}, we indeed recover the conditions \re{CNDS}. \\
The requirements on $U$ and $V$ mentioned in Sec. 3.1 are sufficient to show that the $F$-structures are integrable. The details follow in parallel to the discussion in \cite{Lindstrom:2022nto}.

\subsection{The $(1,1)$ description recovered.}

At the $(1,1)$ level extended susy is governed by additional complex structures. For $(4,4)$ susy there are three extra complex structures for plus  and three for minus directions ${\cal J}^\frak A_{(\al)}$. They form a bi-quaternion structure:
\ber
{\cal J}^\frak A_{(\al)}{\cal J}^\frak B_{(\al)}=-\delta^{\frak A\frak B}+\epsilon^{\frak {ABC}}{\cal J}^\frak C_{(\al)}~.
\eer{}
We now relate the  $(2,2)$ superspace results  to the bi-quaternion structure in $(1,1)$ superspace.

Reducing to $(1,1)$ entails the following split of $(2,2)$ derivatives $\bbD{}$ into $(1,1)$ derivatives $D$ and explicit susy transformation generators $Q$ 
\ber
\bbD{\al}=\half\left(D_\al-iQ_\al\right)~,~~~\bbDB{\al}=\half\left(D_\al+iQ_\al\right)~,
\eer{}
so that the (twisted) chirality constraints become
\ber
Q_\pm\phi^A=J^A_{~B}D_\pm\phi^B~,~~~Q_\pm\chi^{A'}=\pm J^{A'}_{~B'}D_\pm\chi^{B'}
\eer{}
where $J$ has the canonical form $diag(i\one,-i\one)$. When acting on $\varphi$ we then have
\ber
{\cal J}^{(3)}_{(\pm)}=\left( \begin{array}{cccc}
i&0&0&0\cr
0&-i&0&0\cr
0 &0&\pm i&0\cr
0 &0&0&\mp i\end{array}\right)~,
\eer{J3}
where we have labelled the complex structure $3$ to indicate that it is one of the ${\cal J}^\frak A_{(\pm)}$s.
Using \re{J3} we conclude that
\ber
&&\bbD{\pm}\varphi=\half\left(D_\pm-iQ_\pm\right)\varphi=\pi_{(\pm)}D_\pm\varphi\\[1mm]
&&\bbDB{\pm}\varphi=\half\left(D_\pm+iQ_\pm\right)\varphi=\bar\pi_{(\pm)}D_\pm\varphi~,
\eer{}
where 
\ber
\pi_{(\pm)}:=\half\big(\one -i{\cal J}^{(3)}_{(\pm)}\big)~.
\eer{}
To compare the transformations \re{TFS} to the transformations of the $(1,1)$ components, we need to calculate
\ber\nn
&&U^{(+)}\bar\epsilon^+\bbDB{+}\varphi =\bar\epsilon^+U^{(+)} \bar\pi_{(+)}D_+\varphi\\[1mm]
&&=\bar\epsilon^+D_+\left( \begin{array}{c}
\bar \chi\cr
0\cr
-\bar\phi\cr
0\end{array}\right)~,
\eer{TFSS}
all evaluated at the $(1,1)$ level.

Additional supersymmetries  in the  $(1,1)$ formulation lead to an additional two complex structures ${\cal J}^{(1)}$ and ${\cal J}^{(2)}$ that can be read off from the transformations \re{4tfs} (which look identical in $(1,1)$).
On general grounds we expect the transformations to be a linear combinations of ${\cal J}^{(1)}$ and ${\cal J}^{(2)}$ as described in \cite{Lindstrom:2022nto}. Their  form is\footnote{It again checks with the transformations given in \cite{Gates:1984nk}.}
\ber
\delta^{\pm}\varphi+\bar\delta^{\pm}\varphi=\half\left(\left({\cal J}^{(1)}_{(\pm)}+i{\cal J}^{(2)}_{(\pm)}\right)\epsilon^\pm D_\pm\varphi+\left({\cal J}^{(1)}_{(\pm)}-i{\cal J}^{(2)}_{(\pm)}\right) \bar\epsilon^{\pm}D_\pm\varphi\right).
\eer{add}
The plus complex structures we get from \re{4tfs} are
\ber
{\cal J}^{(1)}_{(+)}=\left(\begin{array}{cccc}
0&0&0&1\\
0&0&1&0\\
0&-1&0&0\\
-1&0&0&0\end{array}\right)~,~~~
{\cal J}^{(2)}_{(+)}=\left(\begin{array}{cccc}
0&0&0&i\\
0&0&-i&0\\
0&-i&0&0\\
i&0&0&0\end{array}\right)~,
\eer{}
so that
\ber
\half\left({\cal J}^{(1)}_{(\pm)}-i{\cal J}^{2}_{(\pm)}\right) =\left(\begin{array}{cccc}
0&0&0&1\\
0&0&0&0\\
0&-1&0&0\\
0&0&0&0\end{array}\right)~.
\eer{}
Using these we read off the $\bar\epsilon^+$ transformation from the last term in \re{add}:
\ber
\bar\delta_{(+)}\left( \begin{array}{c}
\phi\cr
\bar\phi\cr
\chi\cr
\bar \chi\end{array}\right)=\left(\begin{array}{cccc}
0&0&0&1\\
0&0&0&0\\
0&-1&0&0\\
0&0&0&0\end{array}\right)\bar\epsilon^+D_+\left( \begin{array}{c}
0\cr
\bar\phi\cr
0 \cr
\bar \chi\end{array}\right)=
\bar\epsilon^+D_+\left( \begin{array}{c}
\bar\chi\cr
0\cr
-\bar\phi \cr
0 \end{array}\right)
\eer{}
Which indeed reproduces the $\bar\epsilon^+$ transformations in \re{4tfs}. 
The $\bar\epsilon^-$ and $\epsilon^\pm$ transformations are similarly reproduced.

\section{Conclusions}
We have illuminated a method for finding extra non manifest supersymmetries directly in $(2,2)$
superspace for $2d$ sigma models describing sectors of generalised K\"ahler geometry. We have done that using the twisted chiral multiplet of \cite{Gates:1984nk} where all steps can be explicitly verified. This corroborates the new results for semi chiral multiplets in \cite{Lindstrom:2022nto} as well as results on complex linear fields presently being investigated.
\vspace {1cm} 

{\bf Acknowledgments}\\

 \noindent I thank   Martin Ro\v cek and Max Hutt
   for discussions and comments. 
  The research  is supported by the Leverhulme Trust through a Visiting Professorship, which is gratefully acknowledged.

\end{document}